\documentclass[aps,prb,preprint,superscriptaddress]{revtex4-1}
\usepackage{graphicx,amsmath,amsfonts}
\begin{document}
\title{Spin dependence of ferroelectric polarization in the double exchange model for manganites}
\author{I. V. Solovyev}
\email{SOLOVYEV.Igor@nims.go.jp}
\affiliation{Computational Materials Science Unit,
National Institute for Materials Science, 1-1 Namiki, Tsukuba,
Ibaraki 305-0044, Japan}
\affiliation{Department of Theoretical Physics and Applied Mathematics, Ural Federal University,
Mira str. 19, 620002 Ekaterinburg, Russia}
\author{S. A. Nikolaev}
\affiliation{Department of Theoretical Physics and Applied Mathematics, Ural Federal University,
Mira str. 19, 620002 Ekaterinburg, Russia}
\date{\today}

\date{\today}
\begin{abstract}
The double exchange (DE) model is systematically applied for studying the coupling between ferroelectric (FE)
and magnetic orders in several prototypical types of multiferroic manganites.
The model itself was constructed for the magnetically active Mn $3d$ bands
in the basis of Wannier functions and include the effect of screened on-site Coulomb interactions
in the Hartree-Fock approximation.
All model parameters were derived from the first-principles
electronic structure calculations.
The essence of our approach for the FE polarization is to use the Berry phase theory,
formulated in terms of
occupied Wannier functions,
and to evaluate the asymmetric spin-dependent change of these functions in the framework of the DE model.
This enables us to
quantify the effect of the magnetic symmetry breaking and
derive several useful expressions for the electronic polarization ${\bf P}$, depending on the relative directions of spins.
The spin-dependence of ${\bf P}$ in the DE model is given by the isotropic correlation functions ${\bf e}_i \cdot {\bf e}_j$
between directions of neighboring spins. Despite formal similarity with the magnetostriction mechanism, the
magnetoelectric (ME) coupling in the proposed DE theory is not related to the magnetically driven FE atomic displacements and
can exist even in compounds with the centrosymmetric crystal structure, if the spacial distribution of
${\bf e}_i \cdot {\bf e}_j$ does not respect the inversion symmetry.
The proposed theory
is applied to the solution of three major problems:
(i) The magnetic-state dependence of ${\bf P}$ in hexagonal
manganites, using YMnO$_3$ as an example;
(ii) The microscopic relationship between canted ferromagnetism and ${\bf P}$ in monoclinic BiMnO$_3$;
(iii) The origin of FE activity in orthorhombic manganites.
Particularly,
we will
show that for an arbitrary noncollinear magnetic structure, propagating along the
orthorhombic $\boldsymbol{b}$ axis and antiferromagnetically coupled along the $\boldsymbol{c}$ axis,
${\bf P}$ is induced by an \textit{inhomogeneous} distribution
of spins and can be obtained by
scaling the one of the E-type antiferromagnetic (AFM) phase with the prefactor depending only on the relative
directions of spins and being the measure of this spin inhomogeneity.
This picture works equally well for the
twofold (HoMnO$_3$) and fourfold (TbMnO$_3$) periodic manganites.
The basic difference is that, even despite some spin canting
of the relativistic origin
and deviation from the collinear E-type AFM aligment,
the twofold periodic magnetic structure remains strongly inhomogeneous,
that leads to large ${\bf P}$. On the contrary,
the fourfold periodic magnetic structure can be viewed
as a moderately distorted homogeneous spin spiral, which corresponds to much weaker ${\bf P}$.
\end{abstract}

\maketitle
\section{\label{sec:Intro} Introduction}

  The multiferroic materials, which simultaneously exhibit a long-range magnetic order and
a spontaneous electric polarization, have attracted a great deal of attention due to their
potential applications in the next-generation electronic devises as well as
the fundamental interest in the origin of magnetoelectric (ME) coupling.\cite{MF_review}
Among them, there is a very important subclass of materials, which are called ``improper multiferroics'',
where the magnetic order does not simply coexist with the spontaneous polarization,
but breaks the inversion symmetry by itself and, thus, becomes primarily responsible for the
ferroelectric (FE) activity. Because of this intrinsic interconnection between polarization and magnetism,
such materials are expected to exhibit a strong ME coupling.

  There is a large number of theoretical studies, which introduce and emphasize the importance of
different mechanisms of the ME coupling, associated with the spin current;\cite{KNB}
nonrelativistic magnetostriction;\cite{SergienkoPRL,Picozzi,Aguilar,Mochizuki}
inverse Dzyalishinskii-Moriya (DM) mechanism, which is another type of magnetostriction, caused by the
relativistic spin-orbit (SO) coupling;\cite{SergienkoPRB} and spin-dependent
$p$-$d$ hybridization.\cite{Arima} Currently, most of these theories have a phenomenological
status, as each of them is typically oriented on the description of properties, observed in
some narrow group of materials. Presumably, the most striking example is the properties of orthorhombic manganites,
which
are typically interpreted from two completely different standpoints: the FE activity in the twofold periodic systems
(such as HoMnO$_3$ and YMnO$_3$) is ascribed solely to the nonrelativistic magnetostriction mechanism, whereas
in materials with longer magnetic periodicity (such as TbMnO$_3$) it is believed
to have a purely relativistic origin, associated with either the spin current or the inverse DM mechanism.

  In this work we continue to develop the double exchange (DE) theory of the ME coupling,\cite{PRB13}
which is oriented on the wide class of multiferroic manganites with different types of the
crystallographic and magnetic structure.

  Manganites play one of the key roles in the materials science engineering of novel multiferroic compounds.
There are two reasons for it: (i) The orbital ordering, which, in the combination with other factors, assists
the antisymmetric charge transfer and the formation of the spontaneous polarization;\cite{Barone,Yamauchi_rew}
(ii) The high spin state of the Mn$^{3+}$ ions, which is driven by intraatomic Hund's coupling
and plays a crucial role in
the magnetic inversion symmetry breaking: since the nonmagnetic state of the Mn$^{3+}$ sites
would lead to a gigantic loss of the intraatomic Hund's energy, in certain magnetic structures
with competing magnetic interactions, it is more favorable
energetically to keep these sites magnetic, but to abandon the inversion symmetry.\cite{remark.1}

  The basic electronic and magnetic properties of manganites are described by the DE model.\cite{DE,Dagotto}
Although the concept of the double exchange was originally introduced for the analysis of the metallic state,
realized in
hole-doped manganites, today it is
understood much more generally -- as a generic property of high spin compounds with the partially filled
majority-spin states. From a mathematical point of view, the `high spin state' means that the intraatomic exchange
splitting between the majority- and minority-spin states is so large that the effect of the latter states
on the considered properties can be
neglected.

  The reorientation of spins in the DE model may lead to a dramatic change of the electronic structure
and even open the band gap.\cite{Dagotto,PRB01,Hotta}
Therefore, in order to understand the behavior of the FE polarization in the DE model,
it is very important to link it to the change of the electronic structure. This can be achieved by using the Berry phase theory
of polarization, which can be reformulated in terms of the occupied Wannier functions in the real space.\cite{KSV,Resta}
For compounds with the centrosymmetric crystal structure, such procedure naturally gives us the electronic polarization,
induced by a noncentrosymmetric magnetic order.

  Hopefully, the electronic structure of insulating manganites
is characterized by another large parameter $\Delta$,
which is the splitting between the occupied and unoccupied states with the same spin. It is
caused by the Jahn-Teller distortion and additionally enhanced by the screened on-site Coulomb repulsion.
This enables us to use the perturbation theory in order to
evaluate the asymmetric spin-dependent change of the Wannier functions in the first order of $1/\Delta$.\cite{PRB13}
This change will automatically gives the spin dependence of the electronic polarization.

  Needless to say that this mechanism of the ME coupling is essentially nonrelativistic one:
The relativistic spin-orbit interaction can play an important role in stabilizing
noncentrosymmetric magnetic structures. However, once it is known, the FE polarization can be described by
the nonrelativistic DE theory. Therefore, depending on the type of the magnetic structure, this
mechanism can produce a large FE polarization.

  In any model analysis, in order to describe the properties of realistic materials and to elucidate the differences
between these materials, it is very important to make a link to the first-principles calculations. In our
case all parameters of the DE model are derived from the first-principles electronic structure calculations, by
constructing some effective Hamiltonians for the magnetically active states in the basis of Wannier functions.\cite{review2008}
Such a procedure typically gives us very reliable description of multiferroic and other properties of
transition-metal oxides at a semi-quantitative level.\cite{PRB11,PRB12,BiMnO3,PRB12hex}

  In our previous work (Ref.~\onlinecite{PRB13}), we have applied such DE model for the analysis of
ferroelectric (FE) activity in one particular type of manganites, crystallizing in the orthorhombic structure and
forming twofold periodic magnetic structure in the ground state. We have argued that by using the DE model
one can indeed successfully rationalize many aspects of the FE activity in this type of systems. Moreover,
even at a quantitative level,
it reproduces results of more general mean-field Hartree-Fock calculations
for the effective model, which were used
as the starting point for the construction of the DE model. These results are also in a good agreement with
available first-principles electronic structure calculations.\cite{Picozzi,Yamauchi,Okuyama}

  In this work we will systematically apply the DE model for the analysis of
the wide class of multiferroic manganites. We will show that the double exchange is indeed
the key microscopic mechanism, which explains the basic aspects of the FE activity, related to the
interplay between magnetic
and crystallographic symmetries
in various types of
manganites.

  The rest of the article is organized as follows. In Sec.~\ref{sec:Method}, we will introduce the DE model
of the ME coupling in multiferroic manganites. In Sec.~\ref{sec:Results}, we will discuss applications of
this model for different types of manganites. Particularly, we will consider
the orthorhombic systems (the space group $Pbnm$), forming twofold and fourfold periodic magnetic structures
in the ground state (Sec.~\ref{subsec:ortho}),
the monoclinic $C2/c$ phase of BiMnO$_3$ (Sec.~\ref{subsec:BiMnO3}),
and hexagonal manganites, crystallizing in the $P6_3cm$ structure,
using YMnO$_3$ as an example (Sec.~\ref{subsec:hexo}).
Finally, in Sec.~\ref{sec:conc}, we will present a summary of our work.

\section{\label{sec:Method} Double exchange model for magnetoelectric coupling}

  Our strategy consists of the following steps:

  (i) We assume that FE and magnetic properties in the ground state of considered systems can be described
reasonably well at the level of the mean-field theories. It can be
the Kohn-Sham density functional theory or its refinements,\cite{WKohn}
which are widely used in the first-principles electronic structure calculations.
In our case, we focus on the behavior of magnetically active $3d$ bands of manganites and replace
the first-principles calculations for this part by the solution of the realistic Hubbard-type model,
which was rigorously constructed using the technique of Wannier functions.\cite{review2008,JPSJ,W_review}
More specifically, we start with the electronic band structure in the local-density approximation (LDA),
construct the Wannier functions for the $3d$ bands, and calculate the matrix elements of the LDA Hamiltonian
in the basis of these Wannier functions. Such a construction gives us the proper one-electron part of the model Hamiltonian.
Then, we calculate the screened on-site Coulomb interactions for the $3d$ bands, using the simplified version of the
constrained random-phase approximation,\cite{Ferdi04} as explained in Ref.~\onlinecite{review2008}.
After that, we solve the effective Hubbard-type model in the mean-field Hartree-Fock (HF) approximation.
The solution gives us the
mean-field Hamiltonian of the form:
\begin{equation}
\hat{H}^{\rm MF}_{ij} = \hat{t}_{ij} + \hat{\cal V}_i \delta_{ij}.
\label{eqn:MF}
\end{equation}
Here, $\hat{t}_{ij} = [ t_{ij}^{mm'} \delta_{ss'} ] $ is the proper one-electron part of the Hubbard model
between sites $i$ and $j$
in the basis of
Wannier orbitals $m$ ($m'$) $=$ $xy$, $yz$, $3z^2$$-$$r^2$, $zx$, and $x^2$$-$$y^2$, and
$\hat{\cal V}_i$ is the self-consistent HF potential at the site $i$, which is constructed from the screened Coulomb interactions
and the density matrix.\cite{review2008}
The Wannier functions themselves were constructed using the
projector-operator technique (Refs.~\onlinecite{review2008} and \onlinecite{W_review})
and the orthonormal linear muffin-tin orbitals (LMTO) (Ref.~\onlinecite{LMTO}) as the trial wave functions.
As the LMTO basis is already well localized, such procedure allows us to generate
well localized Wannier functions.
Therefore, the obtained transfer integrals ($t_{ij}^{mm'}$ for $i \ne j$)
are typically restricted by the nearest neighbors, while other contributions are
substantially smaller. Since the LDA band structure is nonmagnetic and we do not
consider explicitly the SO coupling, the matrix $\hat{t}_{ij}$
does not depend on the spin indices ($s$ and $s'$$=$ $\uparrow$ or $\downarrow$).
Without SO interaction, $\hat{\cal V}_i$ is diagonal with respect to $s$ and $s'$:$$
\hat{\cal V}_i =
\left(
\begin{array}{cc}
\hat{\cal V}_i^\uparrow & 0 \\
0 & \hat{\cal V}_i^\downarrow \\
\end{array}
\right),
$$
where each $\hat{\cal V}_i^{\uparrow,\downarrow}$ is the $5$$\times$$5$ matrix in the orbital subspace.

  (ii) We assume that, to a good approximation, $\hat{\cal V}_i^{\downarrow}$ can be replaced by
$\hat{\cal V}_i^{\downarrow} \approx \hat{\cal V}_i^{\uparrow} + \Delta_{\rm ex}$, where $\Delta_{\rm ex}$
is the averaged exchange splitting between the majority- and minority-spin states.
Due to Hund's interactions,
the splitting $\Delta_{\rm ex}$ is large in manganites. Therefore,
the details of $\hat{\cal V}_i^{\downarrow}$ in the unoccupied part of the spectrum
becomes relatively unimportant on the energy scale of $\Delta_{\rm ex}$.
Moreover, one can consider the limit $\Delta_{\rm ex} \rightarrow \infty$,
and replace $\hat{H}^{\rm MF}_{ij}$ by the DE Hamiltonian:\cite{DE,Dagotto}
\begin{equation}
\hat{H}^{\rm DE}_{ij} = \xi_{ij} \hat{t}_{ij} + \hat{\cal V}_i^\uparrow \delta_{ij},
\label{eqn:DE}
\end{equation}
which operates in the subspace of $\uparrow$-spin states, in the local coordinate frame,
specified by the directions of spins
${\bf e}_i = (\cos \varphi_i \sin \vartheta_i, \sin \varphi_i \sin \vartheta_i, \cos \vartheta_i)$.
The prefactor $\xi_{ij}$ in Eq.~(\ref{eqn:DE}) is given by the well known expression:\cite{Dagotto}
$$
\xi_{ij} = \cos \frac{\vartheta_i}{2} \cos \frac{\vartheta_j}{2} +
\sin \frac{\vartheta_i}{2} \sin \frac{\vartheta_j}{2} e^{-i(\varphi_i - \varphi_j)}.
$$

  (iii) The next step is the calculation of the electronic polarization
using the Berry-phase method.\cite{KSV,Resta} For our purpose, it is convenient to use the
real-space formulation of this method, in terms of the occupied Wannier functions $w_n$.
Then, the electronic polarization is given by:
\begin{equation}
{\bf P} = - \frac{e}{V} \sum_{n = 1}^M
\int {\bf r} \, w_n^2({\bf r}) \, d {\bf r},
\label{eqn:PW}
\end{equation}
where $-$$e$ ($e > 0$) is the electron charge,
$V = L V_0$ is the volume of \textit{magnetic} unit cell (with $V_0$ being the volume
of the crystallographic cell
and $L$ being the number of such cells),
and the summation
$n$ runs over the occupied bands. Alternatively, one can sum up unoccupied bands. This should
give us $-{\bf P}$. It is important to note that we treat ${\bf P}$
as a nonrelativistic quantity, which does not explicitly depend on the SO coupling.
Nevertheless, the latter can still define the directions ${\bf e}_i$ of spins in the ground state.

  The spin dependence of ${\bf P}$ in Eq.~(\ref{eqn:PW}) is accumulated in $w_n({\bf r})$.
Then, for each $n$, one can arbitrarily shift the origin of integration in Eq.~(\ref{eqn:PW}).
Since each $w_n({\bf r})$ is normalized, the shifted integral and the original one
will differ by
some vector, which depends on the shift, but does not depend on the spin variables.
Therefore, since we are interested only in the spin dependence of ${\bf P}$,
we shift the origin of each integral in Eq.~(\ref{eqn:PW}) to the position of that
atomic site for which $w_n$ was constructed and drop all concomitant terms, which do not depend
on the spin degrees of freedom.

\begin{figure}
\begin{center}
\includegraphics[height=6cm]{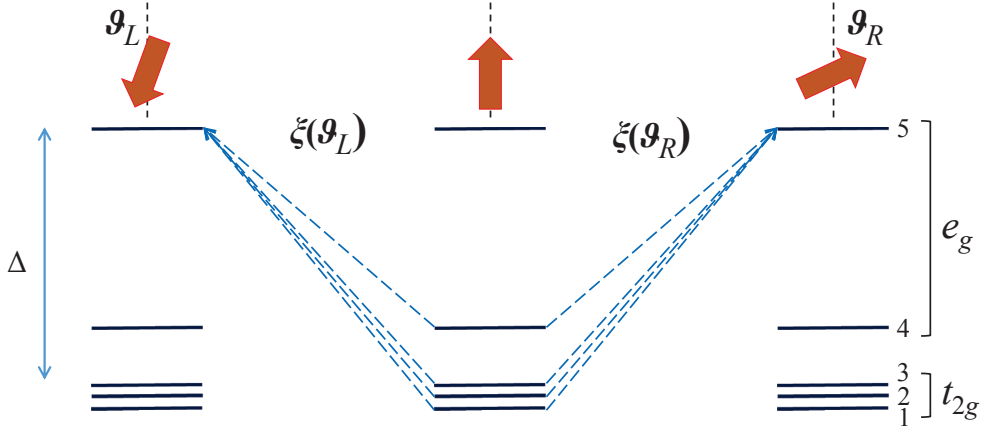}
\end{center}
\caption{\label{fig:idea}(Color online)
Schematic view, explaining the DE model for FE polarization in manganites:
Due to specific magnetic alignment, the spins can form different angles in the bonds lying
on the right ($\vartheta_R$) and on the left ($\vartheta_L$) relative to some center site. In the DE model,
this leads to different scaling of the transfer integrals operating in these two bonds, which is
described by the factors $\xi(\vartheta_R)$ and $\xi(\vartheta_L)$, respectively.
Even if the central site is located in the inversion center, this magnetic alignment breaks the inversion symmetry.
Using the Berry phase theory,\cite{KSV,Resta}
the polarization can be related to
the spin-dependent asymmetric transfer of the weights of the occupied Wannier functions
to the neighboring sites, which are given by transfer integrals between
occupied and unoccupied orbitals (shown by dashed arrows).
Since the unoccupied $e_g$ orbital splits off from the occupied ones by large crystal field
and the Coulomb repulsion (the corresponding splitting is
denoted by $\Delta$),
these transfer integrals can be treated as a perturbation
in the leading order of $1/\Delta$.}
\end{figure}

  (iv) Our next observation is that many manganites exhibit the Jahn-Teller distortion,
which splits the occupied and unoccupied $e_g$ orbitals. This splitting is additionally enhanced by
the on-site Coulomb repulsion. Thus, there is another large parameter $\Delta$,
which characterizes the
electronic structure of manganites. We will define it as
the intraatomic energy splitting between the center of gravity of
occupied manifold, consisting of three $t_{2g}$ and one $e_g$ levels,
and the unoccupied $e_g$ level (see Fig.~\ref{fig:idea}).
The large $\Delta$ allows us to use the perturbation theory for the occupied Wannier functions $w_n$
and evaluate their change in the first order of $1/\Delta$.
Practically, $\Delta$ is obtained form the
diagonalization of the site-diagonal part $(\hat{t}_{ii} + \hat{\cal V}_i^\uparrow)$
of the DE Hamiltonian (\ref{eqn:DE}), which specifies the so-called crystal-field representation
(if there are two types of Mn sites, like in BiMnO$_3$,
we use the averaged value of $\Delta$).
Note that the occupied states in manganites are typically not well separated
from each other and form one broad band.
From this point of view, when we consider the energy splitting, it is more reasonable to use only one energy for all
occupied states and take it in the center of gravity of these states.

  The transfer integrals can be also transformed to the crystal-field representation:
$\hat{t}_{ij} \rightarrow \hat{\mathsf{t}}_{ij}$. Then, one can start with the atomic
limit, where all $w_n$ are fully localized on their atomic sites, and consider the
transfer of weight of $w_n$ to the neighboring sites in the first order
of $\hat{\mathsf{t}}_{ij}/\Delta$. Since the transfers within occupied states
correspond to some unitary transformation of $w_n$, they will not
change the physical properties. Thus, it is sufficient to consider only the
transfer integrals, connecting the occupied and unoccupied orbitals in the crystal-field representation.
As was pointed out above, alternatively, one can consider the change of the unoccupied states and
the transfer integrals from the
unoccupied to occupied orbitals, which is more convenient for our purposes.
Since, at each Mn site, there is only one unoccupied orbital, one can drop the index $n$ in the notations of $w_n$
and replace it by the site index $i$.
Furthermore, we adopt the lattice model and assume that
all weights of $w$ are localized in the lattice points: i.e., if $w_i$ is centered at the site $i$,
its weight can be presented in the form
$$
w_i^2({\bf r}) = \sum_j w_{ij}^2 \, \delta({\bf r} - \Delta \boldsymbol{\tau}_{ji}),
$$
where $\Delta \boldsymbol{\tau}_{ji} = \boldsymbol{R}_j - \boldsymbol{R}_i$ is the position of the site $j$ relative to the site $i$.
Then, in the atomic limit, the weight is accumulated at the central site
($w_{ij}^2 = 1$ and $0$ for $j = i$ and $j \ne i$, respectively). Then, the weights at the neighboring sites can be obtained in
the first order of the perturbation theory for $w_i$ as
\begin{equation}
w_{ij}^2 = \frac{1}{\Delta^2} \sum_{m \le 4} | \mathsf{t}_{ij}^{5m} |^2,
\label{eqn:WFPT}
\end{equation}
where the summation runs over four occupied orbitals.
Since $\hat{t}_{ji} = \hat{t}_{ij}^T$, the weights $w^2_{ij}$ and $w^2_{ji}$
can be obtained from the same matrix of transfer integrals.

  (v) Finally, in the DE model, the transfer integrals $\hat{\mathsf{t}}_{ij}$ should
be additionally modulated by $\xi_{ij}$, which depends on the relative orientation
of spins at the sites $i$ and $j$. Thus, even though the crystal structure itself is centrosymmetric
and the Mn sites are located in the inversion centers (like in orthorhombic manganites,
crystallizing in the $Pbnm$ structure),
the multipliers $\xi_{ij}$ can make some bonds, connecting the central Mn site with its neighbors, inequivalent.
For example, if the right bond $0R$ in Fig.~\ref{fig:idea} is transformed to the left bond $0L$
by the inversion operation, the spin alignment yielding $| \xi(\vartheta_R) | \ne | \xi(\vartheta_L) |$
will make these bonds inequivalent. This will break the inversion symmetry and produce some `dipole',
associated with the site $i$, which will contribute to the electronic polarization as
\begin{equation}
{\bf P}_i = \frac{e}{V} \sum_j  \Delta \boldsymbol{\tau}_{ji}
| \xi_{ij} |^2  w_{ij}^2
\label{eqn:PPT}
\end{equation}
(note that the sign was changed because here we consider the change of the \textit{unoccupied} band and
the electron transfer from the \textit{unoccupied} $e_g$ orbital $5$ to the occupied orbitals $1$-$4$).

  In order to obtain the total polarization, one should sum up all inequivalent dipoles, induced by the
magnetic symmetry breaking. Moreover, it is straightforward to show that
\begin{equation}
| \xi_{ij} |^2 = \frac{1}{2} \left( 1 + {\bf e}_i \cdot {\bf e}_j \right).
\label{eqn:xiasee}
\end{equation}
Therefore, the spin dependence of ${\bf P}_i$ in the DE model
is given by the isotropic correlation functions, ${\bf e}_i \cdot {\bf e}_j$,
between directions of spins. This behavior should also specify the temperature dependence of the
polarization, associated with the spin disorder.

  The spin dependence of ${\bf P}$ has the same form as for the phenomenological magnetostriction mechanism,\cite{Moriya}
which is frequently used for the analysis of magnetoelectric (ME) coupling in manganites.\cite{Aguilar,Mochizuki}
Nevertheless, the new point of our analysis is that this dependence is natural result of the DE physics
and is not necessary related to the magnetically driven FE displacements. Formally speaking,
the proposed DE mechanism can take place even in a centrosymmetric crystal structure
without any magnetostriction, although these two effects can coexist: Once the inversion symmetry is
broken by the magnetic order, there will be magnetostrictive forces, which will move the atoms
away from their centrosymmetric positions. This will activate the magnetostriction mechanism.
In terms of the modern Berry phase theory of polarization,\cite{KSV,Resta} one can say that the DE and magnetostriction mechanism
give rise to, respectively, electronic and ionic parts of the polarization.
According to the first-principles calculations, the electronic contribution in manganites is at least equally important
as the ionic one and cannot be neglected.\cite{Picozzi}

  By summarizing this section: in the DE model, the inversion symmetry is broken by
noncentrosymmetric modulation of the transfer integrals by $\xi_{ij}$.
However, this modulation will also interplay with the crystallographic symmetry of manganites, which is
reflected in the behavior of $\Delta \boldsymbol{\tau}_{ji}$ and $\hat{\mathsf{t}}_{ij}$.
In the next section, we will illustrate how this interplay
will work for different types of manganites, giving rise to finite ME coupling.

\section{\label{sec:Results} Results and Discussions}

\subsection{\label{subsec:ortho} Orthorhombic manganites}

  The connection between FE polarization and magnetic structure of orthorhombic ($Pbnm$) manganites
was discussed in our previous article (Ref.~\onlinecite{PRB13}), also on the level of the DE model.
In this section we present a more general and more comprehensive analysis of the problem.

\begin{figure}
\begin{center}
\includegraphics[height=6cm]{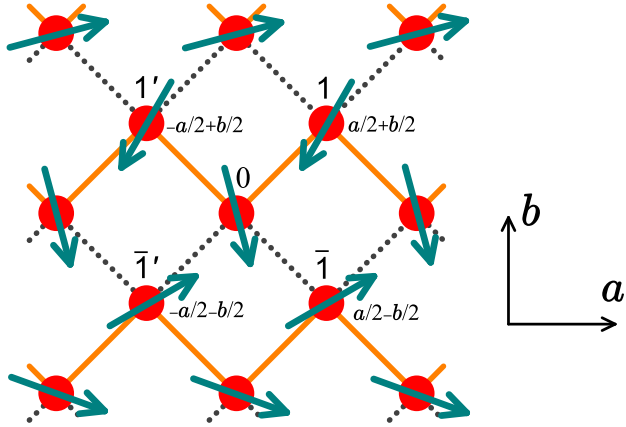}
\end{center}
\caption{\label{fig:OrthoSchem}(Color online)
Schematic view on the crystal and magnetic structures in the $\boldsymbol{ab}$ plane
of orthorhombic manganites. Mn atoms are indicated by circles.
Two types of bonds, which are transformed to themselves by the
symmetry operation $\{ \hat{C}^2_a| \boldsymbol{a}/2$$+$$\boldsymbol{b}/2 \}$, are shown by
solid and broken lines. The magnetic structure is periodic along the orthorhombic $\boldsymbol{a}$ axis
and may have arbitrary periodicity along the $\boldsymbol{b}$ axis.}
\end{figure}

  We assume that the manganites form the perfect antiferromagnetic (AFM) order along the
orthorhombic $\boldsymbol{c}$ axis,
that is indeed consistent with the experimental data.\cite{KimuraNature} Then, in the DE model, all
$\boldsymbol{ab}$ planes become effectively decoupled, and it is sufficient to consider the
single plane. Moreover, it is assumed that the magnetic structure is periodic along the
$\boldsymbol{a}$ axis and may have arbitrary periodicity
along the $\boldsymbol{b}$ axis (see Fig.~\ref{fig:OrthoSchem}),
being again in total agreement with the experimental situation.\cite{Ishiwata}
The most known examples are the twofold periodic E-type AFM structure,
realized in HoMnO$_3$ and YMnO$_3$,\cite{ExpStructureHoMnO3,Ishiwata,Okuyama} and
the nearly fourfold periodic ``spiral'' magnetic structure, realized in TbMnO$_3$.\cite{KimuraNature,KimuraPRB05,ArimaPRB05,Ishiwata}
However, it was also suggested that
both types of magnetic structures are deformed by relativistic interactions and
this deformation has a profound effect on the value of the FE polarization.\cite{PRB11,PRB12}

  The crystal structure in the $\boldsymbol{ab}$ plane can be generated by two symmetry operation:
$\{ \hat{C}^2_a| \boldsymbol{a}/2$$+$$\boldsymbol{b}/2 \}$ (the $180^\circ$ rotation around the
orthorhombic $\boldsymbol{a}$ axis, combined with the translation by $\boldsymbol{a}/2$$+$$\boldsymbol{b}/2$)
and the inversion $\hat{I}$. Moreover, the Mn sites are located
in the inversion centers.

  Then, we take an arbitrary Mn site (`0' in Fig.~\ref{fig:OrthoSchem}) and evaluate its
contribution to the FE polarization, using Eq.~(\ref{eqn:PPT}). In this case, we should sum up
the contributions of four bonds: $0$-$1$, $0$-$1'$, $0$-$\bar{1}'$, and $0$-$\bar{1}$, which correspond to
$\Delta \boldsymbol{\tau}_{ji}=$ $\boldsymbol{a}/2$$+$$\boldsymbol{b}/2$,
$-$$\boldsymbol{a}/2$$+$$\boldsymbol{b}/2$, $-$$\boldsymbol{a}/2$$-$$\boldsymbol{b}/2$, and
$\boldsymbol{a}/2$$-$$\boldsymbol{b}/2$, respectively.
The inversion $\hat{I}$ transforms the bonds $0$-$\bar{1}'$ and $0$-$\bar{1}$ to the bonds $0$-$1$ and $0$-$1'$, respectively,
and the symmetry operation $\{ \hat{C}^2_a| \boldsymbol{a}/2$$+$$\boldsymbol{b}/2 \}$ transforms the
bond $0$-$1'$ to the bond $1$-$0$. Hence, we have $w_{0\bar{1}'} = w_{01}$, $w_{0\bar{1}} = w_{01'}$, and
$w_{01'} = w_{10}$. Moreover, the periodicity of the magnetic structure
along $\boldsymbol{a}$ imposes the following constraints:
$\xi_{01'} = \xi_{01} \equiv \xi_0^+$ and $\xi_{0\bar{1}'} = \xi_{0\bar{1}} \equiv \xi_0^-$, where the
notations $\xi_0^+$ and $\xi_0^-$
stand for the bonds spreading in the positive and negative
directions of the $\boldsymbol{b}$ axis, starting from the site $0$.
Then, the vector ${\bf P}_0$ can be presented in the form:
${\bf P}_0 = \frac{1}{2L}\left\{ |\xi_0^+|^2 - |\xi_0^-|^2 \right\}(P_0^a,P_0^b,0)$, where
\begin{equation}
P_0^a = \frac{ea}{V_0}
\left( w_{01}^2 - w_{10}^2 \right)
\label{eqn:PE}
\end{equation}
is the FE polarization of the E-type AFM phase ($P_0^a \equiv P_E$),\cite{PRB13}
$$
P_0^b = \frac{ea}{V_0}
\left( w_{01}^2 + w_{10}^2 \right).
$$
Since all Mn sites are located strictly in the $\boldsymbol{ab}$ plane and,
in the DE model, all transfer integrals to the neighboring planes are suppressed by the AFM order,
there will be no polarization parallel to the orthorhombic $\boldsymbol{c}$ axis.

  Then, we can repeat this procedure and evaluate the contributions
associated with the neighboring sites $1$ and $\bar{1}$ in the magnetic cell (see Fig.~\ref{fig:OrthoSchem}).
Clearly, half of these contributions will involve the same bonds $0$-$1$, $0$-$1'$, $0$-$\bar{1}'$, and $0$-$\bar{1}$.
Moreover, since the sites $1$ and $\bar{1}$ are obtained from the site $0$ by the
symmetry operation $\{ \hat{C}^2_a| \boldsymbol{a}/2$$+$$\boldsymbol{b}/2 \}$, in the corresponding expressions
for $P_1^a = P_{\bar{1}}^a$ and $P_1^b = P_{\bar{1}}^b$
we will have to interchange $w_{01}$ and $w_{10}$. Furthermore, we note that
$\xi_1^- = \xi_0^+$ and $\xi_{\bar{1}}^+ = \xi_0^-$. Then, one can see that $P_1^b = P_{\bar{1}}^b = P_0^b$.
Therefore, all such contributions
will be canceled out. On the other hand, the contributions parallel to $\boldsymbol{a}$
will satisfy the condition $P_1^a = P_{\bar{1}}^a = -$$P_0^a$.
These contributions can be regrouped so that the total
polarization parallel to $\boldsymbol{a}$ can be presented in the following form:
\begin{equation}
P^a = \frac{1}{L} \sum_{i=0}^{2L-1} (-1)^i |\xi_i^+|^2 P_E,
\label{eqn:Portho1}
\end{equation}
where the summation runs over inequivalent sites of the magnetic unit cell
(note also that the crystallographic cell of
orthorhombic manganites contains two Mn sites in the $\boldsymbol{ab}$ plane).

  Thus, the polarization will be parallel to the $\boldsymbol{a}$ axis. Similar result was obtained in
our previous work,\cite{PRB13} where we considered the magnetic structures, which respect the
symmetry operation $\{ \hat{C}^2_a| \boldsymbol{a}/2$$+$$\boldsymbol{b}/2 \}$.
The present work indicates that this result is more general and does not require any specific
symmetry of the magnetic structure, apart from its periodicity along the $\boldsymbol{a}$ axis and
the AFM coupling along the $\boldsymbol{c}$ axis. Another important point is that
the FE polarization in all magnetic structures can be obtained by scaling the one in
the E-type AFM state. The scaling factor depends on the relative orientation of spins.
Below, we will study it more in details.

  Then, Eq.~(\ref{eqn:xiasee}) can be rewritten for our purposes as
$|\xi_i^+|^2 = \frac{1}{2}\left( 1 + {\bf e}_i \cdot {\bf e}_{i+1} \right)$.
By substituting it in Eq.~(\ref{eqn:Portho1}) and noting that only the direction-dependant part of $|\xi_i^+|^2$
will contribute to $P^a$, one can find that
\begin{equation}
P^a = \frac{1}{2L} \sum_{i=0}^{2L-1} (-1)^i {\bf e}_i \cdot {\bf e}_{i+1} P_E.
\label{eqn:Portho2}
\end{equation}
This formula has the following consequences:

  (i) In the perfect E-type AFM structure, ${\bf e}_i \cdot {\bf e}_{i+1}$ is equal to
$+$$1$ and $-$$1$ for the even and
odd $i$, respectively. However, in the latter case, the minus sign will be additionally changed due to
the prefactor $(-1)^i$.
Therefore, all terms in the sum will be equal to $1$ and we will indeed obtain that $P^a = P_E$.

  One can also consider the deformation of the E-type AFM state, where the odd sublattice is additionally
rotated relative to the even one by the angle $\phi$ (see, e.g., Fig.~3 of Ref.~\onlinecite{Picozzi}).
Such a deformation is caused by the SO interaction
and was obtained in the HF calculations for the realistic Hubbard-type model.\cite{PRB11,PRB12}
In this case, we have ${\bf e}_i \cdot {\bf e}_{i+1} = \pm \cos \phi$
and, therefore,
\begin{equation}
P^a = \cos \phi P_E.
\label{eqn:P2}
\end{equation}
Thus, for small $\phi$, the first correction to $P_E$ appears only in the
second order of $\phi$. This explains that the FE polarization in the E-phase is relatively robust
against the small canting of spins.
However, $P^a$ vanishes in the spin-spiral state, corresponding to $\phi = 90^\circ$.

  (ii) The previous claim appears to be more general and can be reformulated as follows: Any homogeneous arrangement of spins,
which is characterized by the same values of ${\bf e}_i \cdot {\bf e}_{i+1}$ for all bonds in the
$\boldsymbol{ab}$ plane, does not break the inversion symmetry.
All these states have zero electric polarization, that directly follows from Eq.~(\ref{eqn:Portho2}).
Such a situation is realized, for instance, in the ferromagnetic (FM) state or in the
\textit{homogeneous spin-spiral state}, despite a widespread believe.\cite{MF_review} Therefore, in order to obtain
the finite polarization, it is essential to \textit{deform the spin spiral}.
Let us consider such a deformed spin-spiral structure in the
$\boldsymbol{ab}$ plane, for which
${\bf e}_i = \left( \cos \varphi_i, \sin \varphi_i, 0 \right)$ and $\varphi_i = \boldsymbol{q} \cdot \boldsymbol{R}_i + \alpha_i$.
Namely, the phase $\boldsymbol{q} \cdot \boldsymbol{R}_i$ describes the propagation of the homogeneous spin spiral and
the small parameters $\alpha_i$, satisfying the condition $\sum_{i=0}^{2L-1} \alpha_i = 0$,
describe its deformation. The general geometry of the magnetic structure implies that
$\boldsymbol{q} = (0, q, 2 \pi/c)$ and $qbL = 2 \pi n$, where $n$ is an integer. Then, since
${\bf e}_i \cdot {\bf e}_{i+1} = \cos(\varphi_{i+1}$$-$$\varphi_i)$, we will have
\begin{equation}
P^a = \frac{1}{2L} \sum_{i=0}^{2L-1} (-1)^i \cos \left( \frac{\pi n}{L} + \alpha_{i+1} - \alpha_i \right) P_E.
\label{eqn:P4}
\end{equation}
In the first order of $(\alpha_{i+1} - \alpha_i)$, this expression can be further transformed to
$$
P^a \approx -\frac{1}{2L} \sin \left( \frac{\pi n}{L} \right)  \sum_{i=0}^{2L-1} (-1)^i (\alpha_{i+1} - \alpha_i) P_E.
$$
Thus, the spin-spiral inhomogeneity contributes to $P^a$ in the first order of $\{ \alpha_i \}$.
If the phases $\{ \alpha_i \}$ are small, the corresponding polarization is also expected to be small.

  (iii) Eq.~(\ref{eqn:PE}) clearly shows the similarities and differences between
the FE
polarization and interatomic superexchange interactions. Indeed, the expression
$\Delta(w_{01}^2 + w_{10}^2)$ is nothing but the (minus) energy gain caused by the FM alignment of spins
in the bond $0$-$1$,
which contributes to the superexchange interaction.\cite{KugelKhomskii}
Thus, this superexchange interaction is given by the same parameters $w_{01}^2$ and $w_{10}^2$.
The basic difference is that the superexchange interaction is given by the symmetric part of $w_{01}^2$ and $w_{10}^2$,
while the polarization depends on the antisymmetric part.

\begin{figure}
\begin{center}
\includegraphics[height=6cm]{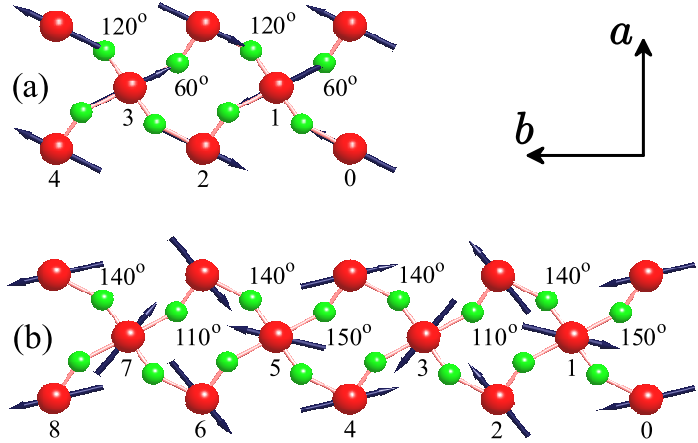}
\end{center}
\caption{\label{fig:HoMnO3}(Color online)
Twofold periodic (a) and fourfold periodic (b) spin structures, as obtained in
the mean-filed Hartree-Fock calculations for the effective Hubbard-type model,
constructed for the $Pbnm$ phase of HoMnO$_3$. The manganese atoms are
indicated by the big red (dark) spheres and the oxygen atoms are indicated by the small
green (grey) spheres. The same numbering of Mn atoms is used in Table~\ref{tab:spiral}.
Other numbers stand for the angles $(\varphi_{i+1}-\varphi_i)$
between spin magnetic moments in the Mn-O-Mn bonds.}
\end{figure}

  Below, we present some numerical estimates, using results of
previous calculations for the magnetic ground state and FE polarization of
orthorhombic manganites.\cite{PRB11,PRB12,PRB13}
Namely, we take the magnetic structures, obtained in the unrestricted HF calculations with the SO coupling, and analyze the
behavior of the FE polarization in terms of the simplified expressions (\ref{eqn:P2}) and (\ref{eqn:P4}),
obtained in the DE model for the deformed magnetic structures of the
AFM E-type and spin-spiral type, respectively.
In the HF calculations, we typically obtain two types of magnetic structures with the
twofold ($L=2$) and fourfold ($L=4$)
periodicity. For HoMnO$_3$, they are shown in Fig.~\ref{fig:HoMnO3}.
Similar behavior was obtained for TbMnO$_3$ and YMnO$_3$.\cite{PRB11,PRB12} The magnetic structures with
larger periodicity typically include the fragments of low periodic structures and the domain wall-defects
(see Fig.~4 of Ref.~\onlinecite{PRB11}).

  As was explained above, the twofold periodic structure can be viewed as the deformed
E-type AFM state. The deformation is caused by the single-ion anisotropy.
For the $Pbnm$ phases,
the angle $\phi$, characterizing this deformation, varies from $63^\circ$, in the case of TbMnO$_3$,\cite{PRB11} till
$60^\circ$ in the case of YMnO$_3$ and HoMnO$_3$ (Ref.~\onlinecite{PRB12} and Fig.~\ref{fig:HoMnO3}).
Then, $P^a$ can be estimated using Eq.~(\ref{eqn:P2}) as $P^a \approx 0.5 P_E$, which is perfectly consistent
with results of the HF calculations. Indeed, the values of $P_E$, obtained in the HF calculations
for the collinear E-type AFM phase without the SO coupling, are $0.96$, $1.09$, and $1.04$ $\mu$C/cm$^2$ for
TbMnO$_3$, HoMnO$_3$, and YMnO$_3$, respectively. The same calculations, but with the SO coupling, yield
$P^a =$ $0.47$, $0.57$, and $0.55$ $\mu$C/cm$^2$ for TbMnO$_3$, HoMnO$_3$, and YMnO$_3$, respectively.
Thus, in all three examples, the above relationship $P^a \approx 0.5 P_E$ works very well and this tendency is nicely explained
by the DE model, where the SO interaction is used in order to obtain the direction of spins
in the ground state, while the FE polarization is calculated as a nonrelativistic quantity
for the given distribution of spins.

  Similar analysis can be done for the fourfold periodic structure, which can be viewed as a deformed
spin spiral, propagating along the $\boldsymbol{b}$ axis. In this case, the spin spiral is stabilized by
isotropic magnetic interactions and deformed by anisotropic and DM interactions.\cite{PRB11}
First, we take the angles $\varphi_i$,
characterizing the directions of spins in the
actual HF calculations with the SO coupling (see Fig.~\ref{fig:HoMnO3}),
and decompose them into the homogeneous ($\boldsymbol{q} \cdot \boldsymbol{R}_i$) and
inhomogeneous ($\alpha_i$) parts. Then, $(qb/2)$ is the averaged value of $( \varphi_{i+1}$$-$$\varphi_i )$
in the magnetic supercell and
$(\alpha_{i+1}$$-$$\alpha_i) = ( \varphi_{i+1}$$-$$\varphi_i ) - qb/2$. For all three compounds,
we obtain $q = 3/4$ (in units of reciprocal lattice translation $\mathrm{g}_b = 2\pi/b$),
which is close to the equilibrium values $q \approx 0.68 \div 0.72$, obtained
in the spin-spiral calculations without the SO coupling.\cite{PRB11}
The parameters $(\alpha_{i+1}$$-$$\alpha_i)$, characterizing the deformation of the spin spiral,
are summarized in Table~\ref{tab:spiral}.
\begin{table}[h!]
\caption{Parameters $\Delta \alpha_i \equiv \alpha_{i+1}$$-$$\alpha_i$, characterizing deformation of the
spin-spiral state (in degrees), as obtained in the unrestricted Hartree-Fock calculations with the spin-orbit coupling
for the $Pbnm$ phase of TbMnO$_3$, HoMnO$_3$, and YMnO$_3$. The atomic positions are explained in Fig.~\ref{fig:HoMnO3}.}
\label{tab:spiral}
\begin{ruledtabular}
\begin{tabular}{cccc}
  Parameters                        &       TbMnO$_3$ &       HoMnO$_3$ &        YMnO$_3$ \\
\hline
$\Delta \alpha_0 = \Delta \alpha_4$ & $\phantom{-}18$ & $\phantom{-}15$ & $\phantom{-}13$ \\
$\Delta \alpha_1 = \Delta \alpha_5$ & $\phantom{-3}7$ & $\phantom{-2}5$ & $\phantom{-2}5$ \\
$\Delta \alpha_2 = \Delta \alpha_6$ &         $-$$32$ &         $-$$25$ &         $-$$23$ \\
$\Delta \alpha_3 = \Delta \alpha_7$ & $\phantom{-3}7$ & $\phantom{-2}5$ & $\phantom{-2}5$ \\
\end{tabular}
\end{ruledtabular}
\end{table}
Then, $P^a$ can be estimated using Eq.~(\ref{eqn:P4}) (note, that $q=3/4$ corresponds to $n=3$), which yields
$P^a/P_E \approx$ $0.115$, $0.081$, and $0.077$ for TbMnO$_3$, HoMnO$_3$ and YMnO$_3$, respectively.
These values are well consistent with results of unrestricted HF calculations
without additional approximations ($P^a/P_E =$ $0.138$, $0.110$, and $0.101$ for TbMnO$_3$, HoMnO$_3$ and YMnO$_3$, respectively).

  Thus, the FE polarization in the ``spin-spiral'' phase is about one order of magnitude smaller than in the
collinear E-phase, in agreement with the experimental data.\cite{Ishiwata} However, this polarization is
caused by the \textit{deformation} of the spin spiral (and not by the spin-spiral alignment itself). Even in the
``spin-spiral'' phase, the FE polarization can be obtained by scaling the one of the E-phase, where the scaling factor
depends only on the relative directions of spins, and all dependencies on the crystal structure itself
are incorporated into $P_E$.

\subsection{\label{subsec:BiMnO3} Monoclinic BiMnO$_3$}

  BiMnO$_3$ is another important compound in the field of multiferroics, and also the
most controversial one. It was regarded as a canonical example of multiferroics, where
the ferroelectricity indeed coexists with the ferromagnetism, but because of two different
mechanisms: the lone pair effect of a nonmagnetic origin was believed to be responsible for the
noncentrosymmetric atomic displacements, which simply coexist with the magnetic properties,
developed in the Mn sublattice.\cite{SeshadriHill}
However, this point of view was basically refuted by subsequent experimental studies
(Ref.~\onlinecite{belik_07}) and electronic structure calculations (Ref.~\onlinecite{spaldin_07}), which suggest
that BiMnO$_3$ should crystallize in the centrosymmetric $C2/c$ structure.
A ``compromised'' point of view was proposed in Ref.~\onlinecite{BiMnO3}, which suggests that BiMnO$_3$
could be an \textit{improper} multiferroic, where the inversion symmetry is broken by some hidden
AFM order. This magnetic inversion breaking gives rise not only to the FE activity,
but also to the DM interactions across the inversion centers, which lead to the
FM canting of spins.\cite{SCLR}

\begin{figure}
\begin{center}
\includegraphics[height=8cm]{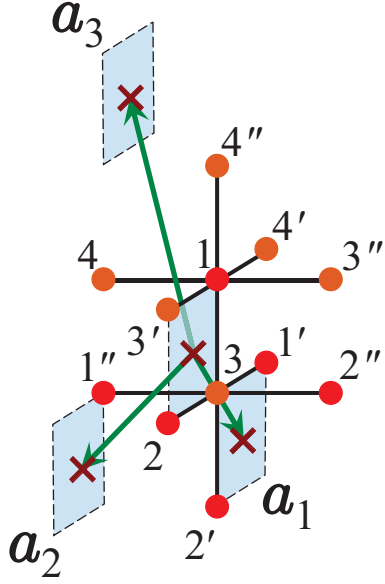}
\end{center}
\caption{\label{fig:C2c}(Color online)
Schematic view on the pseudocubic $C2/c$ structure of BiMnO$_3$.
There are four Mn sites in the primitive cell, which are labeled as $1$, $2$, $3$ and $4$.
These sites form two inequivalent subgroups: ($1$,$2$) and ($3$,$4$), which are shown by different colors.
Other atoms, located in the nearest neighborhood of the inequivalent sites $1$ and $3$,
are obtained by the primitive translations $\boldsymbol{a}_1$, $\boldsymbol{a}_2$, and $\boldsymbol{a}_3$.
The inversion centers (marked by the symbols $\times$) are located in the
centers of the distorted cube face, formed by the atoms $1$, $2$, and two atoms of the type either $3$
or $4$ (for instance $1$-$3$-$2$-$3'$ in the figure).}
\end{figure}

  The monoclinic $C2/c$ phase of BiMnO$_3$ has four formula units (see Fig.\ref{fig:C2c}).
In the following, it is convenient to work with the fractional coordinates, where each vector
$\boldsymbol{v} \equiv (v^1,v^2,v^3)$
is given in
terms of the primitive translations
$\boldsymbol{a}_1 =\frac{1}{2}(a \sin \beta, -b, a \cos \beta)$,
$\boldsymbol{a}_2 =\frac{1}{2}(a \sin \beta, b, a \cos \beta)$, and
$\boldsymbol{a}_3 =(0, 0, c)$
as $\boldsymbol{v} = v^1\boldsymbol{a}_1$$+$$v^2\boldsymbol{a}_2$$+$$v^3\boldsymbol{a}_3$.
Then,
the Mn sites, which are labeled in the figure as $1$, $2$, $3$ and $4$, are located at
$(x,-$$x,1/4)$, $(-$$x,x,-$$1/4)$, $(1/2,0,0)$, and $(0,1/2,1/2)$, respectively.
We use the experimental structure parameters $a = 9.529$ \AA, $b = 5.604$ \AA, $c = 9.848$ \AA,
$\beta = 110.58^\circ$, and $x = 0.21537$, measured at 4~K.\cite{belik_07}
The Mn atoms form two inequivalent groups: ($1$,$2$) and ($3$,$4$). The atoms $1$ and $2$
can be transformed to each other by the inversion operation $\hat{I}$, while the atoms $3$ and $4$
are connected by the symmetry operation $\{ \hat{C}^2_y| \boldsymbol{a}_3/2 \}$.
Therefore, it is sufficient to consider the dipoles around the sites $1$ and $3$.
Similar results for the sites $2$ and $4$ can be obtained by applying the symmetry operations
of the space group $C2/c$. Moreover, the atoms in each of the inequivalent groups are
surrounded by the atoms from another group.

  The crystallographic symmetry can be further lowered by the magnetic order. In this work
we consider the scenario where the magnetic group of BiMnO$_3$ has only one nontrivial symmetry
operation: $\{m_{y}|\textbf{a}_{3}/2\}$
(the mirror reflection $y$$\rightarrow$$-$$y$, combined with the translation by $\textbf{a}_{3}/2$).
This symmetry was indeed obtained in the previous
calculations.\cite{BiMnO3}
Within each inequivalent subgroups of atoms, ($1$,$2$) and ($3$,$4$), $\{m_{y}|\textbf{a}_{3}/2\}$
transforms the Mn sites to each other.
Therefore, if ${\bf e}_1 = (e_1^x,e_1^y,e_1^z)$ is the direction of spin at the site $1$,
the one at the site $2$ will be given by ${\bf e}_2 = (-$$e_1^x,e_1^y,-$$e_1^z)$
(note that ${\bf e}$ is the \textit{axial} vector, and here $x$, $y$, and $z$ denote the directions
in the \textit{monoclinic} frame).
Similar property holds for ${\bf e}_3$ and ${\bf e}_4$.
Thus, $y$-projections of spins are ordered ferromagnetically, while the $x$- and $z$-projections are
ordered antiferromagnetically.
Then, we will have the following symmetry properties:
${\bf e}_1 \cdot {\bf e}_4 = {\bf e}_2 \cdot {\bf e}_3 = e_1^y e_3^y - {\bf e}_1^\perp \cdot {\bf e}_3^\perp$
and ${\bf e}_2 \cdot {\bf e}_4 = {\bf e}_1 \cdot {\bf e}_3$, where ${\bf e}^\perp$ is the AFM
component of spin being perpendicular to $y$. Thus, if $e_i^x = e_i^y = 0$, we obtain the so-called
$\uparrow \downarrow \uparrow \downarrow$ AFM phase, which breaks the inversion symmetry.\cite{BiMnO3}
If $e_i^x = e_i^z = 0$, we deal with the regular FM phase, which preserves the inversion symmetry.

  First, let us consider the contribution of the site $1$ to the FE polarization.
It has six nearest neighbors: $3$, $3'$, $3''$, $4$, $4'$, and $4''$, which are located at
$(1/2,0,0)$, $(-$$1/2,0,0)$, $(1/2,-$$1,0)$, $(0,1/2,1/2)$, $(1,-$$1/2,1/2)$, and $(0,-$$1/2,1/2)$, respectively.
Then, Eq.(\ref{eqn:PPT}) will yield
$$
\begin{aligned}
{\bf P}_1 & =
  | \xi_{13} |^2 \frac{e}{V} \left(
w_{13}^2 \Delta \boldsymbol{\tau}_{31}
 +
w_{13'}^2 \Delta \boldsymbol{\tau}_{3'1}
 +
w_{13''}^2 \Delta \boldsymbol{\tau}_{3''1}
 \right)\\
&+ | \xi_{14} |^2 \frac{e}{V} \left(
w_{14''}^2 \Delta \boldsymbol{\tau}_{4''1}
 +
w_{14}^2 \Delta \boldsymbol{\tau}_{41}
 +
w_{14'}^2 \Delta \boldsymbol{\tau}_{4'1}
\right).
\end{aligned}
$$
The symmetry operation $\{ \hat{C}^2_y| \boldsymbol{a}_3/2 \}$ transforms the site $1$ to itself,
and the sites $3$, $3'$, and $3''$ to the sites $4''$, $4$, and $4'$, respectively. Therefore, we
will have the following properties: $w_{14''}^2 = w_{13}^2$, $w_{14}^2 = w_{13'}^2$, and $w_{14'}^2 = w_{13''}^2$.
Similar expression for ${\bf P}_2$ can be obtained by applying the inversion operation and replacing
$\xi_{13}$ and $\xi_{14}$ by $\xi_{23}$ and $\xi_{24}$, respectively. Then, using the symmetry
properties of $w_{ij}^2$ and $|\xi_{ij}|^2$, together with Eq.~(\ref{eqn:xiasee}),
and noting that
$\Delta \boldsymbol{\tau}_{31} - \Delta \boldsymbol{\tau}_{4''1} = \Delta \boldsymbol{\tau}_{34''} \equiv (1/2,1/2,-$$1/2)$ and
$\Delta \boldsymbol{\tau}_{3'1} - \Delta \boldsymbol{\tau}_{41} = \Delta \boldsymbol{\tau}_{3''1} - \Delta \boldsymbol{\tau}_{4'1} =
\Delta \boldsymbol{\tau}_{3'4} \equiv (-$$1/2,-$$1/2,-$$1/2)$,
one can find that
\begin{equation}
{\bf P}_1 + {\bf P}_2 =
{\bf e}_1^\perp \cdot {\bf e}_3^\perp \frac{e}{V}  \left(
w_{13}^2 \Delta \boldsymbol{\tau}_{34''}
+
[ w_{13'}^2 + w_{13''}^2 ]  \Delta \boldsymbol{\tau}_{3'4}
\right).
\label{eqn:BMOP12}
\end{equation}

  Similar analysis can be performed for the sites $3$ and $4$. For example, the nearest neighbors of the site $3$ are:
$1$, $1'$, $1''$, $2$, $2'$, and $2''$, which are located at $(x,-$$x,1/4)$, $(1$$+$$x,-$$x,1/4)$, $(x,1$$-$$x,1/4)$,
$(-$$x,x,-$$1/4)$, $(1$$-$$x,x,-$$1/4)$, and $(1$$-$$x,x$$-$$1,-$$1/4)$, respectively (see Fig.~\ref{fig:C2c}).
Moreover, the bonds $3$-$1'$ and $3$-$1''$ can be transformed by regular
translations to the bonds $3'$-$1$ and $3''$-$1$, respectively.
The bonds $3$-$2$, $3$-$2'$, and $3$-$2''$ can be transformed to the bonds $3'$-$1$, $3$-$1$, and $3''$-$1$,
respectively, by combining the inversion operation with appropriate translations. Therefore, we will have
the following symmetry properties: $w_{32}^2 = w_{3'1}^2$, $w_{32'}^2 = w_{31}^2$, $w_{32''}^2 = w_{3''1}^2$,
$\Delta \boldsymbol{\tau}_{13} = -$$\Delta \boldsymbol{\tau}_{2'3} = -$$\Delta \boldsymbol{\tau}_{31}$,
$\Delta \boldsymbol{\tau}_{1'3} = -$$\Delta \boldsymbol{\tau}_{23} = -$$\Delta \boldsymbol{\tau}_{3'1}$, and
$\Delta \boldsymbol{\tau}_{1''3} = -$$\Delta \boldsymbol{\tau}_{2''3} = -$$\Delta \boldsymbol{\tau}_{3''1}$,
which yield
$$
{\bf P}_3 =
  \left( | \xi_{32} |^2 - |\xi_{31} |^2 \right) \frac{e}{V} \left(
w_{31}^2 \Delta \boldsymbol{\tau}_{31}
 +
w_{3'1}^2 \Delta \boldsymbol{\tau}_{3'1}
 +
w_{3''1}^2 \Delta \boldsymbol{\tau}_{3''1}
 \right).
$$
Similar expression for the site $4$ is obtained by applying the
symmetry operation $\{ \hat{C}^2_y| \boldsymbol{a}_3/2 \}$, replacing
$\xi_{32}$ and $\xi_{31}$ by $\xi_{42}$ and $\xi_{41}$, respectively, and using the same symmetry properties,
which were used for derivation of Eq.~(\ref{eqn:BMOP12}). Then, one can obtain that
$$
{\bf P}_3 + {\bf P}_4 =
- {\bf e}_1^\perp \cdot {\bf e}_3^\perp \frac{e}{V}  \left(
w_{31}^2 \Delta \boldsymbol{\tau}_{34''}
+
[ w_{3'1}^2 + w_{3''1}^2 ]  \Delta \boldsymbol{\tau}_{3'4}
\right).
$$
Thus, the contribution $({\bf P}_3$$+$${\bf P}_4)$ has the same form as $({\bf P}_1$$+$${\bf P}_2)$,
but with the opposite sign and reversed order of site indices in all of $w_{ij}^2$.

  Then, the total polarization ${\bf P} = \sum_{i=1}^4 {\bf P}_i$
can be obtained by scaling the one in the
$\uparrow \downarrow \uparrow \downarrow$ AFM phase:
\begin{equation}
{\bf P} = {\bf e}_1^\perp \cdot {\bf e}_3^\perp \, {\bf P}_{\uparrow \downarrow \uparrow \downarrow},
\label{eqn:BMOscaling}
\end{equation}
where,
in the fractional coordinate frame:
\begin{equation}
P_{\uparrow \downarrow \uparrow \downarrow}^1 = P_{\uparrow \downarrow \uparrow \downarrow}^2 = -\frac{e}{2V}
\left([w^2_{13'}-w^2_{3'1}] + [w^2_{13''}-w^2_{3''1}] - [w^2_{13}-w^2_{31}]\right),
\label{eqn:BMOP1}
\end{equation}
and
\begin{equation}
P_{\uparrow \downarrow \uparrow \downarrow}^3 =-\frac{e}{2V}
\left([w^2_{13'}-w^2_{3'1}] + [w^2_{13''}-w^2_{3''1}] + [w^2_{13}-w^2_{31}]\right).
\label{eqn:BMOP3}
\end{equation}
It corresponds to
${\bf P}_{\uparrow \downarrow \uparrow \downarrow} =
(P_{\uparrow \downarrow \uparrow \downarrow}^x,0,P_{\uparrow \downarrow \uparrow \downarrow}^z)$
in the cartesian coordinate frame, where
\begin{equation}
P_{\uparrow \downarrow \uparrow \downarrow}^x = \frac{2V}{bc} P_{\uparrow \downarrow \uparrow \downarrow}^1,
\label{eqn:BMOPx}
\end{equation}
\begin{equation}
P_{\uparrow \downarrow \uparrow \downarrow}^z =
2V \cot \beta \, P_{\uparrow \downarrow \uparrow \downarrow}^x + \frac{2 V \sec \beta}{ab} P_{\uparrow \downarrow \uparrow \downarrow}^3,
\label{eqn:BMOPz}
\end{equation}
and the primitive cell volume is $V = \frac{1}{2} abc \sin \beta$.

  Therefore, the properties of the multiferroic phase of BiMnO$_3$ can be rationalized as follows:

    (i) ${\bf P}$ is proportional to the ``correlator'' ${\bf e}_1^\perp \cdot {\bf e}_3^\perp$, constructed from the
AFM components of the neighboring spins.
Therefore, by enforcing the FM alignment of spins along the $y$ axis (e.g., by applying an external magnetic field),
one can decrease ${\bf e}_1^\perp \cdot {\bf e}_3^\perp$ and, therefore, ${\bf P}$. Such a possibility was indeed
investigated in the mean-field HF calculations with the external magnetic field.\cite{BiMnO3}

   (ii) Reversing the AFM components of spins at the sites $3$ and $4$
(${\bf e}_3^\perp \rightarrow -$${\bf e}_3^\perp$), one can flip the direction of polarization
${\bf P} \rightarrow -$${\bf P}$. Particularly, two kinds of
the AFM domains, $\uparrow \downarrow \uparrow \downarrow$ and $\uparrow \downarrow \downarrow \uparrow$,
should have opposite polarization:
${\bf P}_{\uparrow \downarrow \uparrow \downarrow} = -$${\bf P}_{\uparrow \downarrow \downarrow \uparrow}$.
Moreover, by making the AFM projections of spins at the sites $1$ and $3$
to be orthogonal to each other, one can switch off the polarization.

  (iii) Using the values ${\bf P} = (\,-$$0.611,\,0,\,-$$0.042)$ $\mu$C/cm$^2$,
obtained in mean-field HF calculations for the noncollinear magnetic ground state with SO coupling , and
${\bf P}_{\uparrow \downarrow \uparrow \downarrow} = (\,-$$0.790,\,0,\,-$$0.052)$ $\mu$C/cm$^2$,
obtained in the same calculations
for the collinear $\uparrow \downarrow \uparrow \downarrow$ AFM state without SO coupling,\cite{BiMnO3}
one can estimate the ratios
$P^x/P_{\uparrow \downarrow \uparrow \downarrow}^x$ and $P^z/P_{\uparrow \downarrow \uparrow \downarrow}^z$ as
$0.773$ and
$0.808$, respectively. They are well consistent with the value of
${\bf e}_1^\perp \cdot {\bf e}_3^\perp = 0.764$, obtained for the noncollinear magnetic
ground state.\cite{BiMnO3} Thus, the scaling relation (\ref{eqn:BMOscaling}) indeed
works very well and provides a good estimate
for the FE polarization in the noncollinear magnetic ground state of BiMnO$_3$.

  (iv) ${\bf P}_{\uparrow \downarrow \uparrow \downarrow}$ can be estimated using
the Wannier weights, $w^2_{ij}$, collected in Table~\ref{tab:bmo}. The details are summarized in
Supplemental Materials.\cite{SM}
\begin{table}[h!]
\caption{Weights of Wannier functions, $w^2_{ij}$, spreading from the site $i$ to the neighboring site $j$ in BiMnO$_3$.
All parameters are dimensionless, in units of $10^{-3}$. The atomic positions are explained in
Figs.~\ref{fig:C2c} and \ref{fig:BiMnO3OO}.}
\label{tab:bmo}
\begin{ruledtabular}
\begin{tabular}{lcc}
$ij$   & $w^2_{ij}$          & $w^2_{ji}$ \\
\hline
$13$   & $\phantom{1}0.093$  & $13.408$ \\
$13'$  & $\phantom{1}3.801$  & $\phantom{1}5.318$ \\
$13''$ & $13.851$            & $\phantom{1}0.060$ \\
\end{tabular}
\end{ruledtabular}
\end{table}
Particularly, such an analysis allows us to understand why the $z$ component of the polarization is expected to be much
weaker than the $x$ one.

\begin{figure}
\begin{center}
\includegraphics[height=7cm]{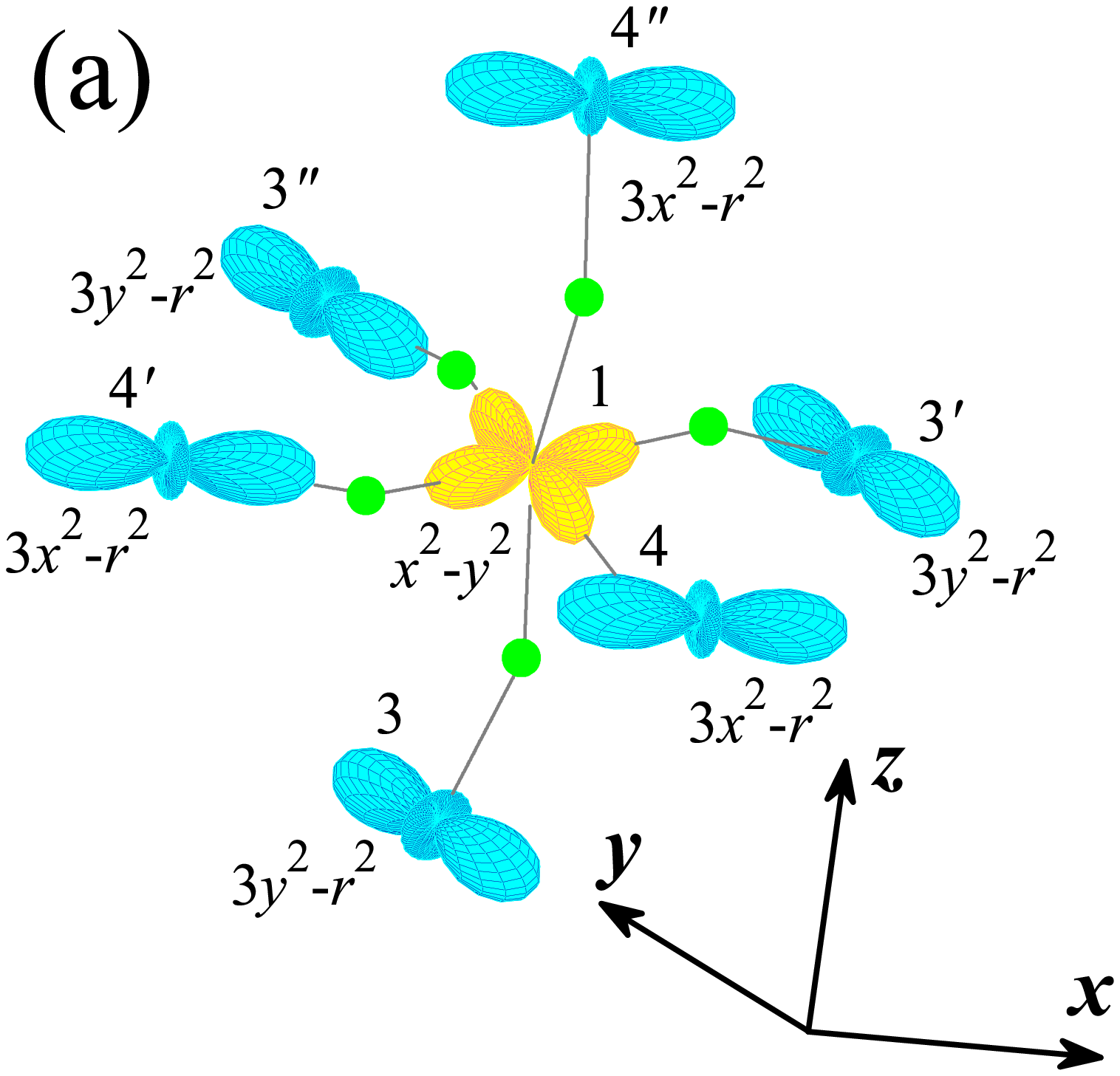}
\includegraphics[height=7cm]{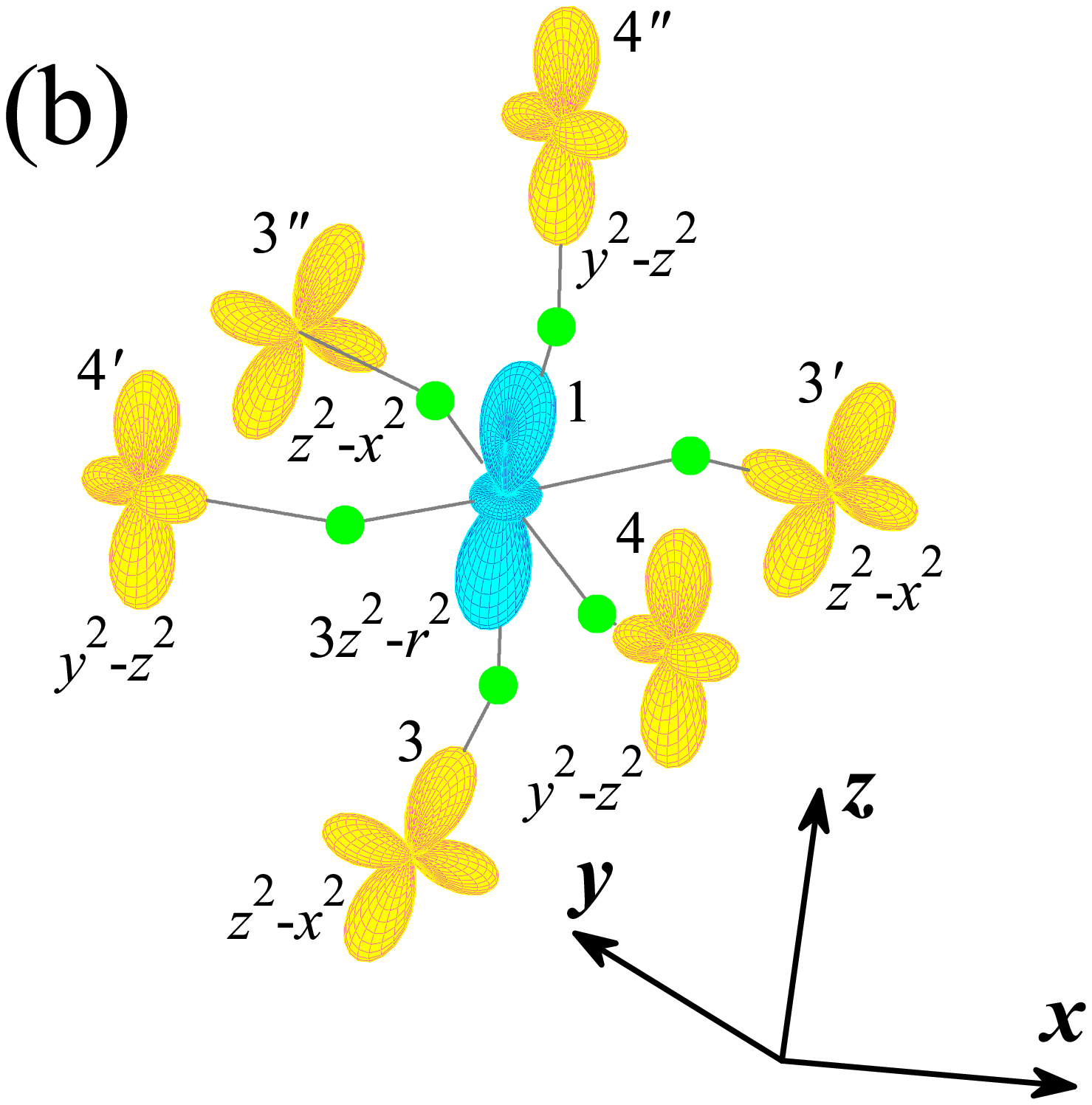}
\end{center}
\caption{\label{fig:BiMnO3OO}(Color online)
Details of orbital ordering in BiMnO$_3$. (a) The unoccupied $e_g$ orbital at the central Mn site $1$ together with
the occupied $e_g$ orbitals at surrounding it Mn sites $3$ and $4$. (b) The occupied $e_g$ orbital at the site $1$ and
the unoccupied $e_g$ orbital at sites $3$ and $4$. The oxygen atoms are indicated by the green (gray) spheres.
The notations of orbitals are only approximate ones and correspond to the perfect cubic environment.
Here, the axes $x$, $y$, and $z$ specify the \textit{pseudocubic} coordinate frame
(should not be confused with the directions $x$, $y$, and $z$ in the \textit{monoclinic} frame, which are used for the
directions of magnetic moment and the symmetry operations of the space group $C2/c$).}
\end{figure}

  First, we note that the contribution of $[w^2_{13'}$$-$$w^2_{3'1}]$ is relatively small.
Moreover, the contributions of $[w^2_{13}$$-$$w^2_{31}]$ and $[w^2_{13''}$$-$$w^2_{3''1}]$ have opposite sign.
Therefore, if for $P_{\uparrow \downarrow \uparrow \downarrow}^3$ there will be a strong cancelation of these two terms,
for $P_{\uparrow \downarrow \uparrow \downarrow}^1$
they will collaborate -- see Eqs.~(\ref{eqn:BMOP1}) and (\ref{eqn:BMOP3}). Thus, $P_{\uparrow \downarrow \uparrow \downarrow}^3$
is expected to be much weaker than $P_{\uparrow \downarrow \uparrow \downarrow}^1$. This behavior is closely related to the
orbital ordering in the bonds $1$-$3$, $1$-$3'$, and $1$-$3''$ (and in the equivalent to them bonds
$1$-$4''$, $1$-$4$, and $1$-$4'$ -- see Fig.~\ref{fig:BiMnO3OO}).
For example, the overlap between the unoccupied $x^2$-$y^2$ orbital at the site $1$ and the occupied $3y^2$-$r^2$ orbital
at the site $3$ is small, that explains the small value of $w^2_{13}$. On the other hand, the overlap between unoccupied
$z^2$-$x^2$ orbital at the site $3$ and occupied $3z^2$-$r^2$ orbital at the site $1$ is much larger, so as the
value of $w^2_{31}$. Therefore, we have $w^2_{13} \ll w^2_{31}$.
The situation in the bond $1$-$3''$ is exactly the opposite and
$w^2_{13''} \gg w^2_{3''1}$. In the bond $1$-$3'$, the overlap between the occupied and unoccupied orbitals is approximately
the same in the both directions and $w^2_{13'} \approx w^2_{3'1}$. If such orbital ordering
were realized in the ideal cubic lattice, one could use the regular Slater-Koster parametrization
for the transfer integrals,\cite{SlaterKoster} which would yield
$[w^2_{13''}$$-$$w^2_{3''1}] = -$$[w^2_{13}$$-$$w^2_{31}]$,
$[w^2_{13'}$$-$$w^2_{3'1}] = 0$, and $P_{\uparrow \downarrow \uparrow \downarrow}^3 = 0$.
Therefore, the second term in Eq.~(\ref{eqn:BMOPz}) will vanish.
Moreover, the first term in Eq.~(\ref{eqn:BMOPz})
will also vanish in the ideal cubic lattice, corresponding to $\beta = 90^\circ$.
Thus, the small value of $P_{\uparrow \downarrow \uparrow \downarrow}^z$ can be regarded as the measure
of deviation from the perfect cubic environment. On the contrary, $P_{\uparrow \downarrow \uparrow \downarrow}^x$
can be finite even in the perfect cubic environment, provided that it supports the specific
type of the orbital ordering, shown in Fig.~\ref{fig:BiMnO3OO}. In this case, $P_{\uparrow \downarrow \uparrow \downarrow}^x$
is given by the simplified expression
$P_{\uparrow \downarrow \uparrow \downarrow}^x = \frac{2e}{bc} [w_{13}^2 - w_{31}^2]$, where
$w_{13}^2$ and $w_{31}^2$ can be obtained using the Slater-Koster parametrization for the ideal cubic lattice.\cite{SlaterKoster}

  Using the values of $w^2_{ij}$ reported in Table~\ref{tab:bmo}, the $x$ component of the polarization can be estimated as
$P_{\uparrow \downarrow \uparrow \downarrow}^x = -$$0.743$ $\mu$C/cm$^2$, which is in excellent agreement with the
value $-$$0.790$ $\mu$C/cm$^2$, obtained in the HF calculations without additional approximations.
However, the agreement for $P_{\uparrow \downarrow \uparrow \downarrow}^z$ is not so good: $0.230$ $\mu$C/cm$^2$
in the present model against $-$$0.052$ $\mu$C/cm$^2$ in the HF calculations.
In the present model analysis, the first term in Eq.~(\ref{eqn:BMOPz}) clearly dominates.
Then, since $\cot \beta < 0$, the sign of the first term in $P_{\uparrow \downarrow \uparrow \downarrow}^z$
should be opposite to $P_{\uparrow \downarrow \uparrow \downarrow}^x$.
This contribution should be compensated by the second term in Eq.~(\ref{eqn:BMOPz}).
However, in the present model analysis, the latter term is found to be small.
Apparently, there are still some contributions, which are missing in the model and which contribute to the
small $z$ component of polarization in the ${\uparrow \downarrow \uparrow \downarrow}$ phase.

\subsection{\label{subsec:hexo} Hexagonal manganites}

\begin{figure}
\begin{center}
\includegraphics[height=6cm]{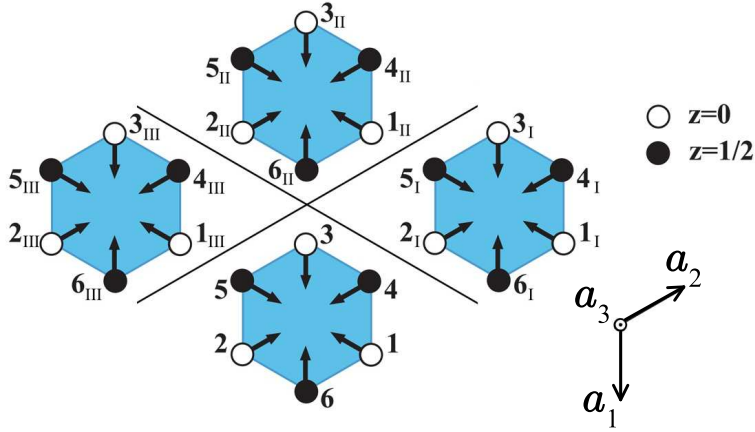}
\end{center}
\caption{\label{fig:hex1}(Color online)
Relative positions of manganese atoms in the hexagonal $P6_{3}cm$ structure. The atoms
located in the $z=0$ and $z=1/2$ planes are indicated by white and black spheres, respectively.
Neighboring unit cells are denoted with Greek numerals. The directions
of spins in the magnetic phase $\Gamma_2$ are shown by arrows.}
\end{figure}

  The hexagonal $P6_{3}cm$ phase of manganites contains six formula units.
By choosing the primitive translations as $\boldsymbol{a}_1 = (0, -$$a, 0)$,
$\boldsymbol{a}_2 = (\frac{\sqrt{3}}{2}a, \frac{1}{2}a, 0)$, and $\boldsymbol{a}_3 = (0, 0, c)$ (see Fig.~\ref{fig:hex1}),
the positions of six Mn atoms, in the fractional coordinates, are
$(x,x,0)$, $(0,-$$x,0)$, $(-$$x,0,0)$, $(0,x,1/2)$, $(-$$x,-$$x,1/2)$ and $(x,0,1/2)$.
The experimental structure parameters for YMnO$_3$ at 10 K, which we consider as an example,
are $a = 6.120$ \AA, $c = 11.407$ \AA, and $x = 0.3423$ (see Supplementary Information of Ref.~\onlinecite{LeeNature}).
All atomic positions can be generated from the first one by applying the
60$^\circ$-degree
rotation around the $z$ axis, combined with the half of the hexagonal translation, $\{C^{6}_{z}|\textbf{c}/2\}$.

\begin{figure}
\begin{center}
\includegraphics[height=6cm]{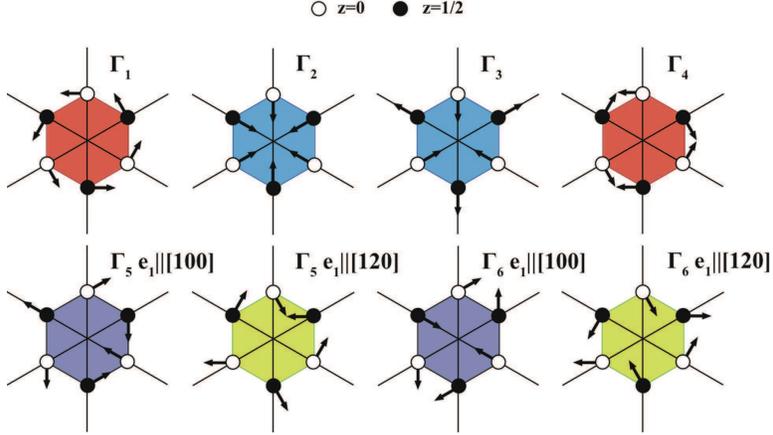}
\end{center}
\caption{\label{fig:hex2}(Color online)
Possible magnetic structures, compatible with the space group $P6_{3}cm$.}
\end{figure}

  There are six possible magnetic structures, which are compatible with the
space group $P6_{3}cm$ (see Fig.~\ref{fig:hex2}).\cite{Munoz,Brown}
Among them, the structures $\Gamma_1$, $\Gamma_4$, $\Gamma_5$ with ${\bf e}_1 \parallel [120]$,
and $\Gamma_6$ with ${\bf e}_1 \parallel [120]$ differ from, respectively,
$\Gamma_2$, $\Gamma_3$, $\Gamma_5$ with ${\bf e}_1 \parallel [100]$,
and $\Gamma_6$ with ${\bf e}_1 \parallel [100]$ by the directions of the easy axes,
which are controlled by the single-ion anisotropy.\cite{PRB12hex}
Thus, the difference is of a relativistic origin.
Since we do not consider explicitly the relativistic effects, we can treat these two groups of states as equivalent.
Moreover, the states $\Gamma_3$ and $\Gamma_5$ differ from, respectively,
$\Gamma_2$ and $\Gamma_6$ by the magnetic alignment in adjacent $xy$ planes:
$\{C^{6}_{z}|\textbf{c}/2\}$ transforms $\Gamma_2$ as the normal symmetry operation, while in $\Gamma_3$,
it is additionally combined with the time-reversal operation $\hat{T}$, which additionally flips the spins
in every second $xy$ plane. For the $\Gamma_5$ and $\Gamma_6$ states, the symmetry operation $\{C^{6}_{z}|\textbf{c}/2\}$
is additionally combined with the $120^\circ$ rotation of spin around the $z$ axis.

  Although these materials have no inversion symmetry and
are ferroelectric irrespectively on their magnetic structure, the transition from the first group of states
($\Gamma_3$ and $\Gamma_5$) to the second one ($\Gamma_2$ and $\Gamma_6$) is characterized by the
finite change of the polarization
(about $-$$120$ $\mu$C/m$^2$, according to the HF calculations for the effective Hubbard-type model).\cite{PRB12hex}
Moreover, the states $\Gamma_2$ and $\Gamma_6$ are weakly ferromagnetic,
giving an interesting possibility for the mutual
control of ferroelectricity and magnetism.
In this section we will elucidate the microscopic origin of the magnetic state dependence of the FE polarization.

  In the following, we assume that all spins lie in the $xy$ plane, and the angles between
neighboring spins in the same plane are fixed
and equal to either $120^\circ$ or $-$$120^\circ$, depending on the type of the magnetic structure.
Then, we consider a continuous transformation from $\Gamma_3$ ($\Gamma_5$)
to $\Gamma_2$ ($\Gamma_6$), where all the spins in every second $xy$ plane are additionally
rotated by the angle $\phi$, varying from $0^\circ$ till $180^\circ$.

  Since relative directions of spins in each $xy$ plane are fixed, there will be no in-plane contributions
to the spin-dependent part of the polarization and we can go directly to the analysis of inter-plane
contributions. First, let us consider the `dipole', associated with the central site $3$ and caused by the
transfer of the weight of the Wannier functions to
the nearest neighbors in the plane, located in the positive direction of $z$.
The corresponding contribution to the spin-dependent part of the polarization can be written as
$$
{\bf P}_3^+ = \frac{e}{2V} \left(
\cos (\phi - \alpha) w^2_{34} \Delta \boldsymbol{\tau}_{43} +
\cos (\phi + \alpha) w^2_{35} \Delta \boldsymbol{\tau}_{53} +
\cos \phi w^2_{36_{\rm II}} \Delta \boldsymbol{\tau}_{6_{\rm II}3}
\right)
$$
(see Figs.~\ref{fig:hex1} and Fig.~\ref{fig:hex2} for the notations of atomic sites
and the relative directions of spin moments, respectively), where $\alpha =$ $120^\circ$ ($-$$120^\circ$)
for $\Gamma_3$ ($\Gamma_5$). Moreover, due to the mirror reflection
$x \rightarrow -$$x$, which is one of the symmetry operations of the space group $P6_{3}cm$,
the bonds $3$-$4$ and $3$-$5$ are equivalent. Therefore, $w^2_{34} = w^2_{35}$ and
the above expression can be further rearranged as
$$
{\bf P}_3^+ = \frac{e}{2V} \left(
\cos \phi \cos \alpha w^2_{34} (\Delta \boldsymbol{\tau}_{43} + \Delta \boldsymbol{\tau}_{53})
+ \sin \phi \sin \alpha w^2_{34} \Delta \boldsymbol{\tau}_{45}
+ \cos \phi w^2_{36_{\rm II}} \Delta \boldsymbol{\tau}_{6_{\rm II}3}
\right),
$$
where, in the fractional coordinates,
$\Delta \boldsymbol{\tau}_{43}$$+$$\Delta \boldsymbol{\tau}_{53} = (x,0,1)$,
$\Delta \boldsymbol{\tau}_{45} = (x,2x,0)$, and
$\Delta \boldsymbol{\tau}_{6_{\rm II}3} = (2x$$-$$1,0,1/2)$.
Similar expressions for other sites can be obtained by applying the symmetry operation
$\{C^{6}_{z}|\textbf{c}/2\}$ to ${\bf P}_3^+$. Then, it is clear that, due to the symmetry,
all $xy$ contributions to the total ${\bf P}^+$, obtained after the summation over six Mn sites in the
primitive cell, will be canceled out, and ${\bf P}^+$ will be parallel to $z$. It is given by the
following expression
$$
{\bf P}^+ = \frac{3e}{2V} \cos \phi \left(
w^2_{36_{\rm II}} - w^2_{34}
\right) \boldsymbol{a}_3,
$$
which holds for both transitions: from $\Gamma_3$ to $\Gamma_2$ and from $\Gamma_5$ to $\Gamma_6$.
Similar expression for ${\bf P}^-$, caused by the transfer of the Wannier weight to
the nearest neighbor sites $\bar{4}$ and $\bar{6}_{\rm II}$, located in the negative direction of $z$,
is obtained by replacing $\boldsymbol{a}_3$ by $-$$\boldsymbol{a}_3$. Moreover, due to the symmetry operation
$\{C^{6}_{z}|\textbf{c}/2\}$, the Wannier weights obey the following properties:
$w^2_{3\bar{6}_{\rm II}} = w^2_{6_{\rm II}3}$ and $w^2_{3\bar{4}} = w^2_{43}$. Altogether, this leads to
$$
{\bf P}^- = -\frac{3e}{2V} \cos \phi \left(
w^2_{6_{\rm II}3} - w^2_{43}
\right) \boldsymbol{a}_3.
$$

  Thus, the dependence of the polarization on the relative directions of spins between neighboring planes
obeys the simplest $\cos \phi$ law. The nearest-neighbors (NN) contribution to the polarization change
$\Delta {\bf P} = {\bf P}(\Gamma_{2,6}) - {\bf P}(\Gamma_{3,5})$,
associated with the transition from
$\Gamma_3$ and $\Gamma_5$ ($\phi = 0$) to, respectively, $\Gamma_2$ and $\Gamma_6$ ($\phi = 180^\circ$),
is given by
$$
\Delta {\bf P}_{NN} = \frac{3e}{V} \left(
w^2_{34} - w^2_{43} - w^2_{36_{\rm II}} + w^2_{6_{\rm II}3}
\right) \boldsymbol{a}_3.
$$
Similar expression, associated with the transfer of the Wannier weights to the next-nearest-neighbor (NNN) sites in the
planes $z = \pm$$1/2$ is obtained by replacing the sites $4$ and $6_{\rm II}$ by the sites $4_{\rm III}$ and $6$,
respectively (see Fig.~\ref{fig:hex1}).

  The Wannier weights, $w^2_{ij}$, obtained for YMnO$_3$ are collected in Table~\ref{tab:ymohex}.
Details can be found in Supplemental Materials.\cite{SM}
\begin{table}[h!]
\caption{Weights of Wannier functions, $w^2_{ij}$, spreading from the site $i$ to the site $j$
in the hexagonal YMnO$_3$.
All parameters are dimensionless, in units of $10^{-4}$. The atomic positions are explained in Fig.~\ref{fig:hex1}.
The first two lines show the data for the inequivalent nearest-neighbor (NN) bonds and the second two lines --
to the next-nearest-neighbor (NNN) bonds between the planes.}
\label{tab:ymohex}
\begin{ruledtabular}
\begin{tabular}{lccc}
$ij$           & type & $w^2_{ij}$   & $w^2_{ji}$ \\
\hline
$34$           & NN   & $0.741$      & $0.414$    \\
$36_{\rm II}$  & NN   & $0.985$      & $0.601$    \\
$34_{\rm III}$ & NNN  & $0.005$      & $0.789$    \\
$36$           & NNN  & $1.068$      & $0.047$    \\
\end{tabular}
\end{ruledtabular}
\end{table}
Using these parameters, the nearest-neighbor contribution to $\Delta P$ (parallel yo $z$) can be estimated as
$-$$4$ $\mu$C/m$^2$, which is small due to the strong cancelation between two inequivalent types of bonds.
The contribution of next-nearest neighbors is $-$$134$ $\mu$C/m$^2$, which is consistent with
the value $-$$120$ $\mu$C/m$^2$, obtained in the HF calculations without additional
approximations.\cite{PRB12hex}
Thus, the value of $\Delta P$ is mainly determined by the Wannier transfer between next-nearest neighbors
in adjacent plane. This is consistent with other magnetic properties of hexagonal manganites.
For instance,
the type of the magnetic coupling between the planes
is also controlled by the NNN interactions, while the contribution of the
nearest neighbors is small or can be of the opposite sign.\cite{PRB12hex}

\section{\label{sec:conc}Summary and Conclusions}

  In this work, we extend the DE theory of the ME coupling and systematically apply it
to the wide class of multiferroic manganites, exhibiting different crystallographic and magnetic structures.
For all considered materials, we are able to
present a transparent physical picture of
how the FE polarization is induced and controlled by the magnetic order.
This picture is based on the DE theory, which was formulated for the effective low-energy model,
derived from the first-principles calculations.
Our basic idea is that for the analysis of electronic properties of manganites one can use
two physical limits.
The first one is the DE limit, which means that the
intraatomic exchange splitting between the majority-
and minority-spin states, driven by Hund's interactions, is so large that the
contribution of the latter states to the electronic polarization can be neglected. The second one is the limit of
large intraatomic splitting $\Delta$ between the occupied and unoccupied orbitals with the majority spin,
which is driven by the Jahn-Teller distortion and the screened on-site Coulomb repulsion.
The second limit allows us to use the perturbation theory for the occupied Wannier functions in the
first order of $1/\Delta$, which can be incorporated in the general Berry-phase theory of polarization.
Thus,
the electronic polarization can be described in term of the asymmetric transfer of the weights of the
Wannier functions to the neighboring sites, which, in the DE model, are additionally modulated by the spin-dependent factors.
The DE model allows to greatly simplify the analysis of the polarization
and present it in the transparent form, explaining the interplay between the crystallographic and magnetic structures.
Moreover, it provides the simple analytical
dependence of the FE polarization on the relative directions of spins.
As expected for the nonrelativistic theory of the FE polarization, this dependence is given by the
`isotropic correlators' ${\bf e}_i \cdot {\bf e}_j$.
It has the same form as for the phenomenological magnetostriction mechanism. Nevertheless, this is
a new mechanism, which is not directly related to the magnetostriction and can take place without
magnetostriction, even in a centrosymmetric crystal structure.

  First, we have systematically applied
the DE theory to the orthorhombic $Pbnm$ manganites and generalized
results of our previous work (Ref.~\onlinecite{PRB13}).
Our present result is valid for any noncollinear magnetic structure, of an arbitrary periodicity, propagating
along the orthorhombic $\boldsymbol{b}$ axis and antiferromagnetically coupled along the $\boldsymbol{c}$ axis.
For this type of magnetic structures, we have argued that the FE polarization should be parallel to the
orthorhombic $\boldsymbol{a}$ axis and can be obtained by scaling the one of the collinear E-type AFM state
with the scaling factor depending exclusively on the relative direction of spins.
The FE polarization vanishes in the homogeneous spin-spiral state, which
preserves the inversion symmetry of the DE Hamiltonian. Therefore, the only possibility to obtain the
finite polarization is to deform the homogeneous spin spiral and to produce some
inhomogeneity in the distribution of spins. In multiferroic manganites,
such a deformation is caused by the relativistic SO interaction. This picture works equally well for
manganites with the twofold and fourfold periodic magnetic structures, which typically attributed to
HoMnO$_3$ and TbMnO$_3$, respectively.
The basic difference is that, even despite some spin canting and deviation from the collinear E-type AFM alignment,
the twofold periodic magnetic structure remains strongly inhomogeneous,
that leads to the large polarization. On the contrary,
the fourfold periodic magnetic structure can be viewed
as a distorted homogenous spin spiral.
Therefore, if the distortion is small, the polarization is also small.

  Next, we have studied the microscopic origin of the FE polarization, caused by the magnetic inversion
symmetry breaking in the $C2/c$ phase of BiMnO$_{3}$. The uniqueness of this situation is that the magnetic
ground state of BiMnO$_3$ contains both AFM $\uparrow \downarrow \uparrow \downarrow$ component, which
breaks the inversion symmetry, and the FM magnetization, caused by the canting of spins. Thus, this is a rare case, where the
FE polarization indeed coexists with the FM magnetization, that is very important
from the viewpoint of the mutual control of ferroelectricity and magnetism. According to the mean-field HF calculations,
the AFM magnetization lies in the monoclinic $zx$ plane, while the FM one is parallel to the $y$ axis.\cite{BiMnO3}
We have modeled this magnetic structure in our DE analysis in order to find a quantitative relationship
between the FM magnetization and the FE polarization. As expected, these two quantities `anticorrelate' with each other:
by enforcing the FM magnetization, one can decreases the polarization. The latter can be obtained by scaling
the one of the collinear $\uparrow \downarrow \uparrow \downarrow$ AFM state. The scaling factor is given again
by the correlation function between directions of neighboring spins, but since the FM moments do not contribute
to the polarization, this correlation function includes only AFM components of the magnetization.
The polarization in this case lies in the $zx$ plane (so as the AFM magnetization). Moreover, the $x$ component of
the polarization is substantially larger than the $z$ one. We have found this behavior to be closely related
to the orbital ordering, realized in BiMnO$_{3}$: while the $x$ component is very robust and can be expected
even in the
perfect cubic lattice (provided that it supports the particular
type of the orbital ordering, realized in BiMnO$_3$), the weak $z$
component is the measure of deviation from the perfect cubic environment, which crucially depends on the details
of the monoclinic distortion.

  Finally, we have explained the origin of the ME coupling, associated with the reversal of spins in every second
$xy$ plane of hexagonal manganites. Such a reversal can be indeed expected in realistic materials if one can
induce the change of the magnetic structure from $\Gamma_2$ to $\Gamma_3$ (or between any two types of the
magnetic structures, in which the magnetic moments in the neighboring planes are transformed, respectively, by the
native symmetry operations of the $P6_{3}cm$ space group and by the same symmetry operations, combined with the
time reversal). Although the $P6_{3}cm$ space group has no inversion symmetry and, therefore, the
system is expected to be ferroelectric, irrespectively on the magnetic order, this change of the magnetic
structure produced a finite change of the FE polarization. We have derived an analytical expression for
the spin-dependent part of the polarization and evaluated different contributions to it, associated with the transfer of the weights
of the Wannier functions to different groups of sites in the adjacent planes. We have found that
the main contribution comes from next-nearest neighbors, while the ones from the nearest sites are small
due to the strong cancelation, which occurs between two inequivalent types of bonds.

  In conclusion, the DE mechanism of the ME coupling plays a very important role in physics of multiferroic manganites
and explains many basics aspects of the FE activity in these systems on a unified ground.
Thus, this is the
key microscopic mechanisms, which should be considered
in the analysis of multiferroic properties of manganites and related compounds.

\textit{Acknowledgements}.
This work is partly supported by the grant of Russian Science Foundation (project No. 14-12-00306).


\begin{thebibliography}{99}
\bibitem{remark.1}
The magnetic structure of many multiferroic manganites is
organized in such a way that there is certain group of Mn sites, which are transformed to themselves by the regular inversion.
This typically contradicts to the symmetry properties of some other group of Mn atoms, which, due to specific form of
interatomic magnetic interactions, should be transformed to themselves by the inversion combined with the time reversal operation.
In order to satisfy simultaneously these two symmetry constraints, one of atomic groups should become nonmagnetic.
However, it would correspond to a gigantic loss of intraatomic Hund's energy. Therefore, it is more
favorable energetically to preserve the high spin state of Mn$^{3+}$, but to break the inversion symmetry.
\end{thebibliography}
\end{document}